\documentclass{moriond}

\usepackage{amssymb}
\def\Journal#1#2#3#4{{#1} {\bf #2}, #3 (#4)}

\def\PRL{\em Phys. Rev. Lett.}

\def\be{\begin{equation}}
\def\ee{\end{equation}}
\def\bea{\begin{eqnarray}}
\def\eea{\end{eqnarray}}

\newcommand*{\doi}[2]{\href{http://dx.doi.org/#1}{#2}}

\newcommand{\Photo}{}

\begin{document}
\vspace*{4cm}
\title{TESTING THE GRAVITATIONAL THEORY WITH SHORT-PERIOD STARS AROUND OUR GALACTIC CENTER}

\author{A. HEES$^{1}$, A.~M.~GHEZ$^{1}$, T. DO$^{1}$, J.~R.~LU$^{3}$, M.~R.~MORRIS$^{1}$,  E.~E.~BECKLIN$^{1}$, G.~WITZEL$^{1}$, A.~BOEHLE$^{1}$, S.~CHAPPELL$^{1}$, Z.~CHEN$^1$, D.~CHU$^{1}$, A.~CIURLO$^1$, A.~DEHGHANFAR$^{1}$, E.~GALLEGO-CANO$^2$,  A.~GAUTAM$^1$, S.~JIA$^3$, K.~KOSMO$^{1}$, G.~D.~MARTINEZ$^{1}$, K.~MATTHEWS$^{4}$, S.~NAOZ$^{1}$, S.~SAKAI$^{1}$, R.~SCH\"ODEL$^{2}$}

\address{$^{1}$ Department of Physics and Astronomy, University of California, Los Angeles, CA 90095, USA \\
$^2$ Instituto de Astrof\'isica de Andaluc\'ia (CSIC), Glorieta de la Astronom\'ia S/N, 18008 Granada, Spain\\
$^{3}$ Astronomy Department, University of California, Berkeley, CA 94720, USA \\
$^4$ Division of Physics, Mathematics, and Astronomy, California Institute of Technology, MC 301-17, Pasadena, CA 91125, USA
}

\maketitle\abstracts{Motion of short-period stars orbiting the supermassive black hole in our Galactic Center has been monitored for more than 20 years. These observations are currently offering a new way to test the gravitational theory in an unexplored regime: in a strong gravitational field, around a supermassive black hole. In this proceeding, we present three results: (i) a constraint on a hypothetical fifth force obtained by using 19 years of observations of the two best measured short-period stars S0-2 and S0-38 ; (ii) an upper limit on the secular advance of the argument of the periastron for the star S0-2 ; (iii) a sensitivity analysis showing that  the relativistic redshift of S0-2 will be measured after its closest approach to the black hole in 2018. }

\section{Introduction}

\begin{figure}[htb]
	\begin{minipage}{0.4\linewidth}
		\centerline{\includegraphics[width=0.99\linewidth]{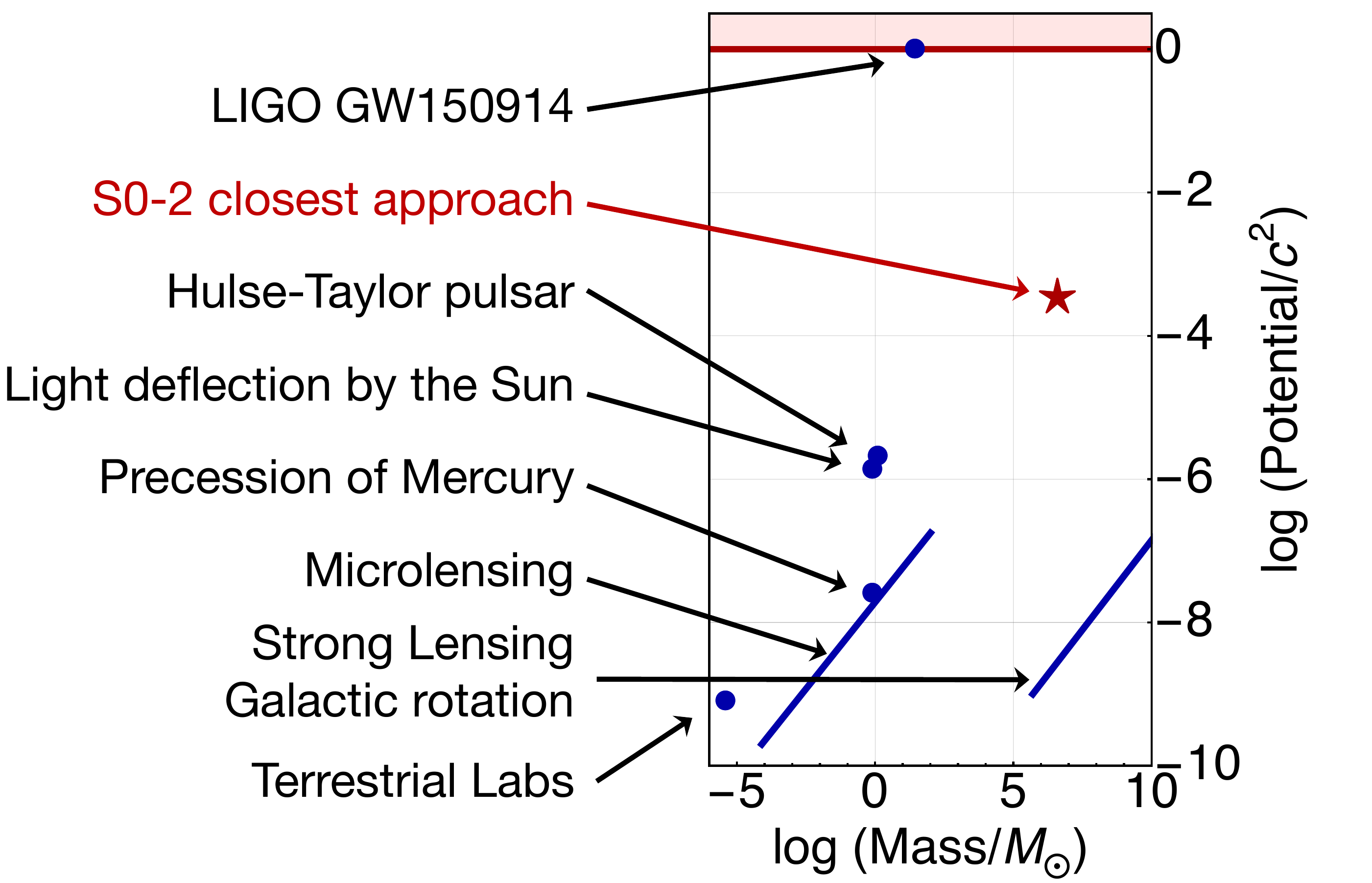}}
	\end{minipage}
\hfill
	\begin{minipage}{0.58\linewidth}
		{\includegraphics[width=0.99\linewidth]{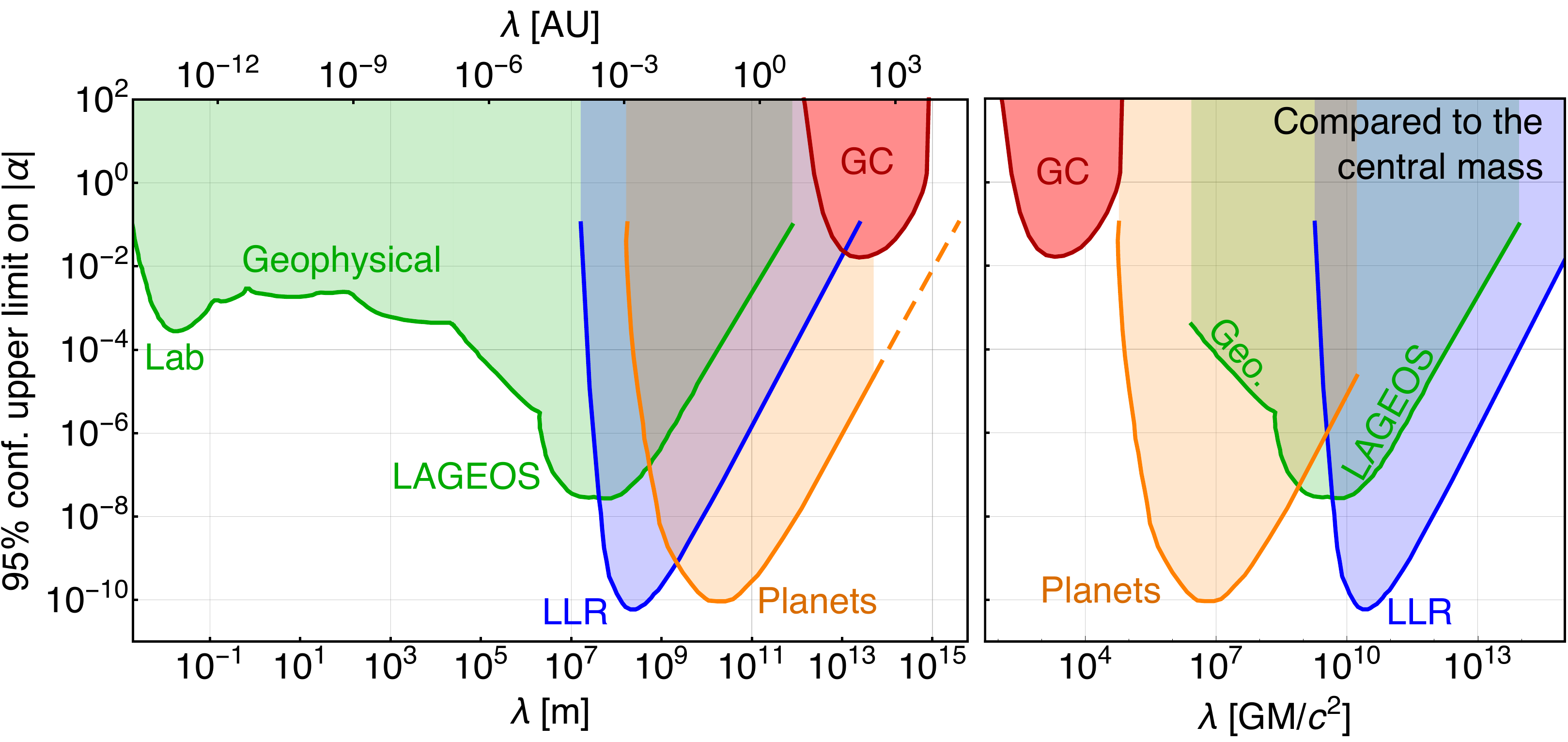}}
	\end{minipage}
\caption[]{Left: The gravitational potential probed by different tests of gravitation against the mass of the central body that generates gravity in these tests. Short-period stars, such as S0-2, around our Galactic Center explore a new region in this parameter space. Figure inspired by Psaltis~\cite{psaltis} (see also~Baker {\it et al}\cite{baker}). \\ Right: 95\% confidence upper limit on the fifth force parameter $\left|\alpha\right|$ as a function of $\lambda$. The shaded regions are excluded by various experiments. Our analysis is represented by the red shaded area (GC)~\cite{hees} while the other curves are from Fig. 31 of Konopliv {\it et al}~\cite{konopliv}. The dashed curve is a reasonable extrapolation based on Solar System results~\cite{konopliv}. Left panel: the horizontal axis is $\lambda$ in meters or in AU. Right panel: the horizontal axis is $\lambda$ expressed in term of the gravitational radius of the central mass that generates gravitation in the different experiments.}
\label{fig:first}
\end{figure}

Testing General Relativity (GR) and constraining alternative gravitational theories has been a long-standing endeavor in the scientific community. The primary motivation is the development of a quantum theory of gravitation, of a theory that would unify all fundamental interactions and by models of dark matter and dark energy. While GR is extremely well tested in the Solar System and with binary pulsars (e.g. Will~\cite{will}), observations of stars orbiting the supermassive black hole (SMBH) at the center of our Galaxy offer the opportunity to probe gravity in a strong field regime~\cite{rubilar,zucker} as depicted on the left of Fig.~\ref{fig:first}.

The motion of short-period stars around our Galactic Center (GC) has been monitored for more than 20 years by two experiments: one carried out at the Keck Observatory~\cite{ghez,boehle} and the other with the Very Large Telescope~\cite{genzel,gillessen17}. These observations have provided proof for existence of a SMBH in our GC~\cite{ghez,genzel} and have been used to determine the distance to our GC with a 2\% relative accuracy~\cite{boehle,gillessen17}. 

In  this proceeding, we show that we are currently entering an era where  observations of short-period stars around our GC can be used to probe fundamental physics. We report a constraint on a hypothetical fifth force obtained using observations of the two best measured stars in our GC. In addition, we present an upper limit on the secular advance of the argument of the periastron for the star S0-2 which can be used to constrain various theoretical and astrophysical scenarios in our GC. Finally, we present a sensitivity analysis that shows that S0-2's closest approach in 2018 will enable the first measurement of relativistic effects with S-stars.

\section{Constraint on a hypothetical fifth force}
The fifth force formalism considers a modification from Newtonian gravity in which the gravitational potential takes the form of a Yukawa potential
\begin{equation}
	U=\frac{GM}{r}\left[1+\alpha e^{-r /\lambda}\right]  \, ,
\end{equation}
where $G$ is the Newton's constant, $M$ is the mass of the central body, $r$ is the distance to the central body, $\alpha$ is the strength of the fifth interaction and $\lambda$ is its length scale. This phenomenological framework is motivated by several theoretical scenarios such as: theories with additional interactions, massive gravity, massive tensor-scalar theories, higher dimensional theories, etc~\cite{fischbach,adelberger}. This phenomenology has motivated many experiments at a wide variety of scales~\cite{adelberger}: in the lab, with geophysical measurements, with satellite and lunar laser ranging and with planetary ephemerides. In addition, it has been shown that a fifth force would impact the short-period star S0-2~\cite{borka}.

We used 19 years of observations of the two stars with the highest orbital phase coverage S0-2 and S0-38 to constrain a hypothetical fifth force in our GC~\cite{hees}. The dataset used is fully described in Boehle {\it et al}~\cite{boehle}. The model used in our orbital fit includes 21 parameters consisting of 6 orbital parameters for each star and of 9 global parameters: the fifth force parameter $\alpha$, the SMBH $GM$, the amount of extended mass enclosed within 0.011 parsec, the distance to our GC, the positions and velocities of the SMBH. The positions and velocities of the SMBH are important to take into account imperfections in the construction of our reference frame. In addition, in order to estimate the systematics uncertainties arising in the construction of the absolute reference frame, a Jackknife  resampling method is used~\cite{boehle,hees}. No deviation from Newtonian gravity is found and upper limit on the fifth force strength $\alpha$ has been derived and is presented on the right of Fig.~\ref{fig:first}. Our best constraint is at the level of $\lambda\sim 150$ astronomical units, with a 95\% confidence upper limit of $\left|\alpha\right|<0.016$. These constraints are complementary to the ones obtained in the Solar System in the sense that they are probing space-time in a new region of the parameter space (see left panel of Fig.~\ref{fig:first}): around a SMBH, which is conceptually different from Solar System tests where the space-time curvature is generated by weakly gravitating bodies and in a much higher gravitational potential. This is highlighted in the right panel from Fig.~\ref{fig:first} where $\lambda$ is expressed in term of the gravitational radius of the central body.  In addition, some alternative theories of gravity exhibit screening mechanisms which may screened the deviations from GR in the Solar System. In this context, searches for deviations from GR in other environments are important. 

\section{Upper limit on the  advance of the argument of periastron of the star S0-2}
Most alternative theories of gravitation predict a different secular drift of the argument of periastron $\omega$ than GR. In addition, several astrophysical scenarios produce a similar effect. Therefore, we derived an upper limit on a linear drift of the argument of the periastron $\dot \omega$ for the star S0-2~\cite{hees} using the 19 years of observations described in Boehle {\it et al}~\cite{boehle}. Our orbital fit includes 7 parameters for each star: the 6 standard orbital parameters and the secular drift of the argument of periastron $\dot \omega$. The global parameters included in our fit consist of the SMBH $GM$, the distance to our GC and the positions and velocities of the SMBH (important to take into account imperfections in the construction of our global reference frame). As a result of our analysis including the Jackknnife resampling to estimate systematics uncertainties, we obtained an upper limit on a linear drift of the argument of periastron for S0-2 given by
\begin{equation}
	\left|\dot\omega_\textrm{{\tiny S0-2}}\right|<1.7\times 10^{-3} \textrm{ rad/yr} \qquad \textrm{at 95\% C.L.}\, .
\end{equation}
This limit is one order of magnitude larger than the relativistic advance of the periastron $\dot\omega_\textrm{\tiny GR}=1.6\times 10^{-4}$ rad/yr for S0-2. Nevertheless, this limit can be used to derive preliminary constraints on various theoretical and astrophysical scenarios in our GC. A similar, although not as stringent, limit can be obtained for the star S0-38 and is given by $\left|\dot \omega_\textrm{\tiny S0-38}\right|\lesssim 7.6\times 10^{-3}$ rad/yr.

\section{Measurement of the relativistic redshift}
\begin{figure}[htb]
\begin{minipage}{0.5\linewidth}
\centerline{\includegraphics[width=0.9\linewidth]{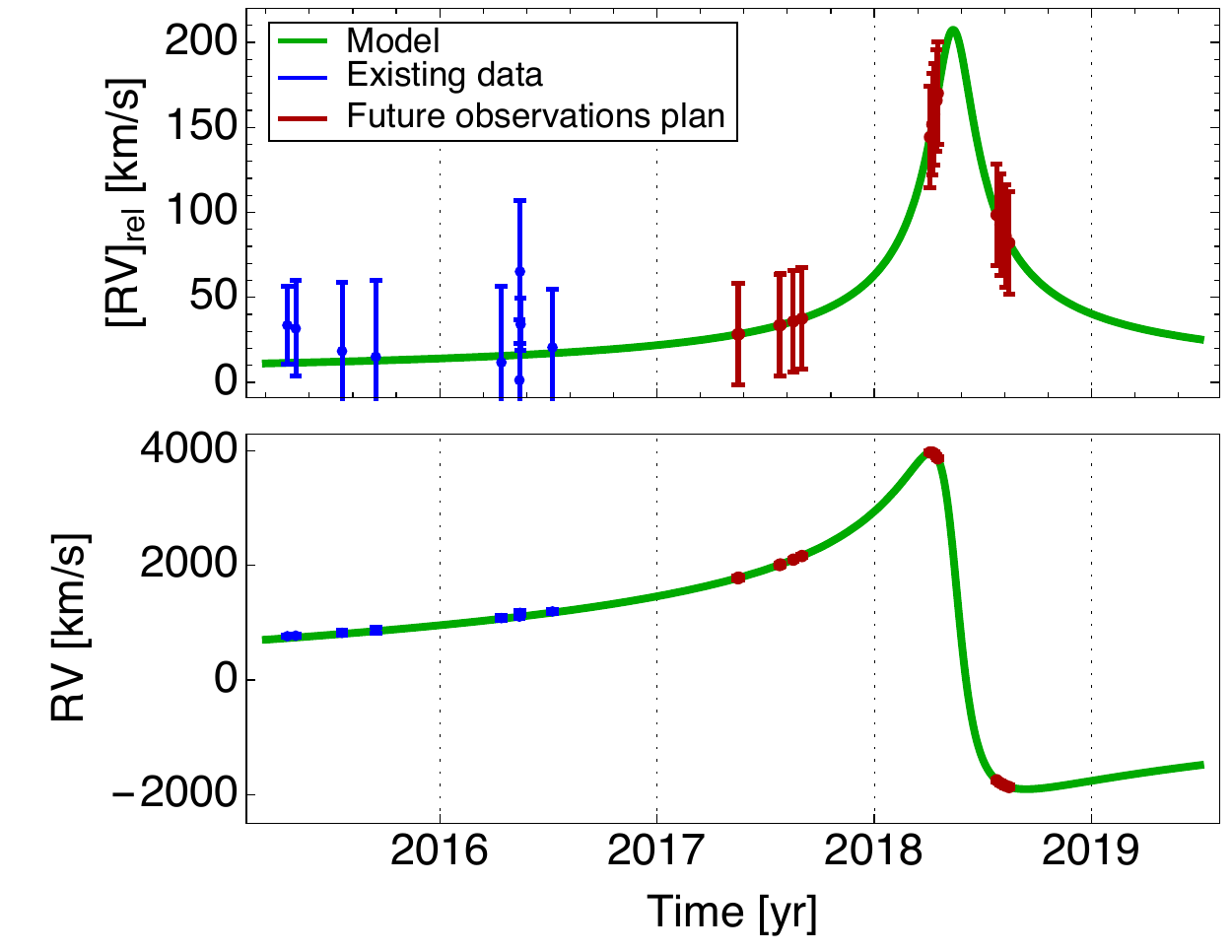}}
\end{minipage}
\hfill
\begin{minipage}{0.5\linewidth}
\centerline{\includegraphics[width=0.75\linewidth]{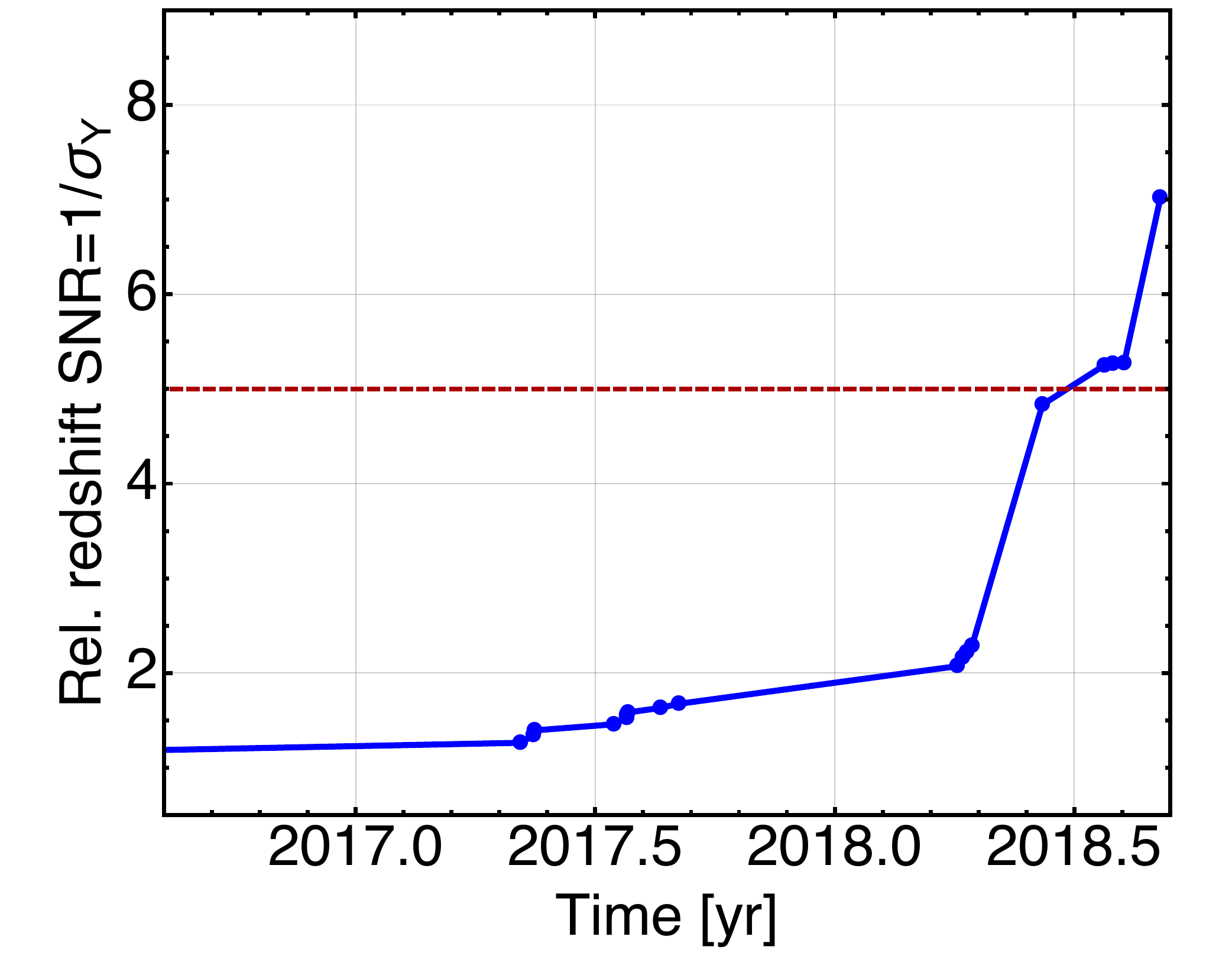}}
\end{minipage}
\caption[]{Left:  S0-2's radial velocity for the total RV (bottom panel) and the relativistic contribution (top panel) from Eq.~\ref{eq:redshift}. The blue points correspond to existing data from Keck and VLT~\cite{gillessen17}. The red points correspond to the an observation plan for 2017 and 2018 optimally designed to detect the relativistic redshift. The optimal observations are the ones corresponding to the turning points of the RV curve.\\
Right: evolution of the signal-to-noise ratio of the relativistic redshift measurement with future observations. The blue points correspond to the future observations.}
\label{fig:RV}
\end{figure}

In 2018, the star S0-2 will reach its closest approach to the SMBH. As anticipated by many~\cite{rubilar,zucker}, this event will allow us to measure the first post-Newtonian effect around a SMBH: the relativistic redshift that impacts S0-2's radial velocity. The contribution of the first order relativistic redshift on the radial velocity is given by
\begin{equation}\label{eq:redshift}
	\left[RV\right]_\textrm{\tiny rel}=\frac{v^2}{2c}+\frac{GM}{rc}  \, ,
\end{equation}
where $r/v$ are the norm of the position and velocity of the star with respect to the SMBH and $c$ the speed of light. The first term is a contribution due to special relativity while the second term corresponds to the gravitational redshift. The relativistic redshift contribution to S0-2's radial velocity reaches 200 km/s as shown on the left of Fig.~\ref{fig:RV}. One way to measure the relativistic redshift is to model the total radial velocity as $	RV=\left[RV\right]_\textrm{\tiny Newton}+\Upsilon \left[RV\right]_\textrm{\tiny rel} \, ,$
where $\left[RV\right]_\textrm{\tiny Newton}$ is the standard Newtonian radial velocity and $\Upsilon$ is a dimensionless parameter whose value is equal to 1 in GR. The idea is to fit $\Upsilon$ simultaneously with the other parameters in the orbital fit: a value significantly different from 0 but compatible with 1 would be evidence of a successful detection of the relativistic redshift while a value significantly different from 1 would indicate a deviation from GR. The current observations of S0-2 are compatible with GR and exclude the Newtonian model at $1.2\sigma$.

 In order to maximize the use of telescope time in 2018, we developed an adaptive scheduling tool that determines the optimal observation nights in order to measure the relativistic redshift, as shown in Fig.~\ref{fig:RV} (left). The optimal observation epochs for a detection of the redshift correspond to the turning points of the radial velocity curve (i.e. the extremal points, see left of Fig.~\ref{fig:RV}). The right of Fig.~\ref{fig:RV} shows the evolution of the signal-to-noise ratio (defined as $1/\sigma_\Upsilon$) of the relativistic redshift measurement with the additional observations. The optimal observations ensure a successful detection of the relativistic redshift above 5$\sigma$ at the end of the 2018 observation campaign.

\section{Conclusion}
In this proceeding, we show that we are currently entering an era where observations of short-period stars in our GC can be used to probe fundamental physics. We presented a constraint on a hypothetical fifth force and an upper limit on the secular drift of S0-2's argument of periastron obtained using 19 years of observations~\cite{hees}. Finally, we have shown that a carefully designed observations plan for 2018 will lead to a successful measurement of the relativistic redshift of the star S0-2. In the longer term, tests of GR using short-period stars are expected to be complementary to other types of observations that will probe the space-time around the SMBH at the center of our Galaxy, as for example with the Event Horizon Telescope~\cite{psaltis16}.

\section*{Acknowledgments}
AH thanks the organizers for financial support to attend the conference. Support for this work was provided by NSF grant AST-1412615, the Heising-Simon Foundation, the Levine-Leichtman Family Foundation, the Galactic Center Board of Advisors, and Janet Marott for her support of research on S0-38 through the Galactic Center stellar patron program.

\end{document}